\documentstyle[aps,prd,multicol]{revtex}
\tighten

\def\Lrule{\vspace*{-0.2in}\noindent\vrule width3.4in height.2pt
  depth.2pt \vrule depth0em height.5em}
\def\Rrule{\vspace{-0.1in}\hfill\vrule depth.5em height0pt \vrule
  width3.4in height.2pt depth.2pt\vspace*{-0.1in}}

\begin{document}

\title{Remarks on the canonical quantization of noncommutative
theories}

\author{R. Amorim$^a$ and J. Barcelos-Neto$^b$}

\address{\mbox{}\\
Instituto de F\'{\i}sica\\
Universidade Federal do Rio de Janeiro\\
RJ 21945-970 - Caixa Postal 68528 - Brasil}
\date{\today}

\maketitle
\begin{abstract}
\hfill{\small\bf Abstract\hspace*{1.7em}}\hfill\smallskip
\par
\noindent
Free noncommutative fields constitute a natural and interesting
example of constrained theories with higher derivatives. The
quantization methods involving constraints in the higher derivative
formalism can be nicely applied to these systems. We study real and
complex free noncommutative scalar fields where momenta have an
infinite number of terms. We show that these expressions can be summed
in a closed way and lead to a set of Dirac brackets which matches the
usual corresponding brackets of the commutative case.

\end{abstract}

\smallskip\mbox{}

\begin{multicols}{2}
\section{Introduction}
\renewcommand{\theequation}{1.\arabic{equation}}

Recently, there have been a great deal of interest in noncommutative
fields. This interest started when it was noted that noncommutative
spaces naturally arise in perturbative string theory with a constant
background magnetic field in the presence of $D$-branes. In this
limit, the dynamics of the $D$-brane can be described by a
noncommutative gauge theory \cite{Witten}. Besides their origin in
strings and branes, noncommutative field theories are a very
interesting subject by their own rights \cite{Seiberg}. They have been
extensively studied under several approaches \cite{Sheikh,Gomis}. To
obtain the noncommutative version of a field theory one essentially
replaces the usual product of fields by the Moyal product
\cite{Witten,Seiberg}, which leads to an infinite number of spacetime
derivatives over the fields. It can be directly verified that the
Moyal product does not alter quadratic terms in the action, provided
boundary terms are discarded. In this way, the noncommutativity does
not affect the equations of motion for free fields. However, on the
other hand, we know that momenta can be obtained as surface terms of a
hypersurface orthogonal to the time direction \cite{Landau}. This
means that momenta are different in the versions with and without
Moyal products. In fact, momenta in the version with Moyal products
have an infinite number of terms.

\medskip
Hence, noncommutative field theories provide us with an interesting
and non
academic example involving higher derivatives where the quantization
rules for such systems can be nicely applied. We emphasize that this
is a very peculiar situation. Usually, nontrivial examples of systems
with higher derivatives are just academical and plagued with ghosts
and nonunitarity problems. In fact, one can say that these systems
have never constituted a confident test for the nonconventional
quantization procedure involving higher derivatives, meanly in the
cases where there are constraints. Free noncommutative theories, even
though can be described without the Moyal product, are then an
interesting theoretical laboratory for using the higher derivative
formalism with constraints. This is precisely the purpose of our
paper. We are going to study the quantization of the free
noncommutative scalar theory without discarding the infinite higher
derivative terms of the Lagrangian. We shall see that, regardless the
completely different expressions of the momenta, the canonical
quantization can be consistently developed in terms of a constraint
formalism.

\medskip
Our paper is organized as follows. In Sec. II we present a simple
example involving the usual free scalar theory (without Moyal product)
but writing the action in terms of higher derivatives. The purpose of
considering this simple example is to present some particularities of
the higher derivative quantization method we are going to use in the
next sections. In Sec. III we deal with free noncommutative real
scalar fields and in Sec. IV we consider the complex case, where the
momentum expressions are still more evolved. We left Sec. V for our
conclusions.

\section{A simple example}
\renewcommand{\theequation}{2.\arabic{equation}}
\setcounter{equation}{0}

Let us consider a simple example of a free scalar field involving
higher derivatives, which is given by the action

\begin{equation}
S=-\,\frac{1}{2}\int d^4x\,\phi\,\Box\phi
\label{2.1}
\end{equation}

\noindent
Of course, disregarding boundary terms at infinity we may rewrite this
action as $S=\frac{1}{2}\int d^4x\,\partial_\mu\phi\partial^\mu\phi$,
from which one obtains the momentum $\pi=\dot\phi$. The canonical
quantization is obtained by transformating the Poisson bracket
$\{\phi(\vec x,t),\pi(\vec y,t)\}=\delta(\vec x-\vec y)$ in the
commutator of field operators $[\phi(\vec x,t),\pi(\vec y,t)]=
i\,\delta(\vec x-\vec y)$. This leads to the commutator $[\phi(\vec
x,t),\dot\phi(\vec y,t)]=i\,\delta(\vec x-\vec y)$ which together with
the equation of motion $\Box\,\phi=0$ permit us to calculate the
propagator and develop all the process of quantization. However, let
us keep the action as given by (\ref{2.1}). For systems with higher
derivatives, velocities are unconventionally assumed to be independent
coordinates\cite{Musik,Barc1}. In the particular case of action
(\ref{2.1}), the independent coordinates are $\phi$ and $\dot\phi$. It
is accepted that there is a conjugate momentum for each of them. These
can be obtained by fixing the  variation of fields and velocities at
just one of the extreme times, say $\delta\phi(\vec x,t_0)=0=
\delta\dot\phi(\vec x,t_0)$ and keeping the other extreme free
\cite{Landau}. The momenta conjugate to $\phi$ and $\dot\phi$ are the
coefficients of $\delta\phi$ and $\delta\dot\phi$ respectively, taken
at time $t$ over a hipersurface orthogonal to $t$.

\medskip
Considering the variation of the action with just the extreme in
$t_0$ kept fixed we have\cite{Barc2}

\begin{eqnarray}
\delta S&=&-\frac{1}{2}\int_{t_0}^tdt\int d^3\vec x\,
\bigl(\delta\phi\,\Box\phi+\phi\Box\delta\phi\bigr)
\nonumber\\
&=&\int_{t_0}^tdt\int d^3\vec x\,\Bigl[-\delta\phi\,\Box\phi
+\frac{1}{2}\partial^\mu\bigl(\partial_\mu\phi\delta\phi
-\phi\partial_\mu\delta\phi\bigr)\Bigr]
\nonumber\\
&=&-\int_{t_0}^tdt\int d^3\vec x\,\delta\phi\,\Box\phi
+\frac{1}{2}\int d^3\vec x\,\bigl(\dot\phi\,\delta\phi
-\phi\,\delta\dot\phi\bigr)
\nonumber\\
&=&\frac{1}{2}\int d^3\vec x\,\bigl(\dot\phi\,\delta\phi
-\phi\,\delta\dot\phi\bigr)
\label{2.2}
\end{eqnarray}

\noindent
where in the last step it was used the on-shell condition $\Box\,\phi=
0$. The quantities $\delta\phi$ and $\delta\dot\phi$ that appear in
(\ref{2.2}) are both taken at time $t$. From this boundary term we
identify the conjugate momenta to $\phi$ and $\dot\phi$, which are the
coefficients of $\delta\phi$ and $\delta\dot\phi$ respectively. We
denote these momenta by

\begin{eqnarray}
\pi&=&\frac{1}{2}\,\dot\phi
\label{2.3}\\
\pi^{(1)}&=&-\frac{1}{2}\,\phi
\label{2.4}
\end{eqnarray}

\noindent
We observe that even though the equation of motion is the same as the
corresponding one of the case without higher derivatives, the momenta
expressions are not.

\medskip
Since now $\phi$ and $\dot\phi$ are independent coordinates, both
expressions above are constraints \cite{Dirac}. This means
that the commutators cannot be inferred from the fundamental Poisson
brackets, namely

\begin{eqnarray}
\{\phi(\vec x,t),\pi(\vec y,t)\}&=&\delta(\vec x-\vec y)
\nonumber\\
\{\dot\phi(\vec x,t),\pi^{(1)}(\vec y,t)\}&=&\delta(\vec x-\vec y)
\label{2.5}
\end{eqnarray}

\noindent
but from the Dirac ones. It is interesting to notice that the
Poisson bracket $\{\phi(\vec x,t),\dot\phi(\vec y,t)\}$ is zero. For
the constraints given by (\ref{2.3}) and(\ref{2.4}) we obtain the
Dirac brackets \cite{Barc2}

\begin{eqnarray}
\{\phi(\vec x),\pi(\vec y)\}_D
&=&\frac{1}{2}\delta(\vec x-\vec y)
\nonumber\\
\{\dot\phi(\vec x),\pi^{(1)}(\vec y)\}_D
&=&\frac{1}{2}\delta(\vec x-\vec y)
\label{2.6}
\end{eqnarray}

\noindent
from which we infer the commutators

\begin{eqnarray}
[\phi(\vec x),\pi(\vec y)]
&=&\frac{i}{2}\delta(\vec x-\vec y)
\nonumber\\
{}[\dot\phi(\vec x),\pi^{(1)}(\vec y)]
&=&\frac{i}{2}\delta(\vec x-\vec y)
\label{2.7}
\end{eqnarray}

\noindent
Using the expressions of the momenta given by (\ref{2.3}) and
(\ref{2.4}) we observe that both relations above lead to the
expected commutator $[\phi(\vec x),\dot\phi(\vec y)]=i\,\delta(\vec
x-\vec y)$, which means that the quantization does not change, as it
should be, even the boundary terms are contributing with different
expressions for the momentum.

\section{Noncommutative case}
\renewcommand{\theequation}{3.\arabic{equation}}
\setcounter{equation}{0}

Let us consider the action

\begin{equation}
S=\frac{1}{2}\int_{t_0}^tdt\int d^3\vec x\,
\partial_\mu\phi\star\partial^\mu\phi
\label{3.1}
\end{equation}

\noindent
where $\star$ is the notation for the Moyal product, whose definition
for two general fields $\phi_1$ and $\phi_2$ reads

\begin{equation}
\phi_1(x)\star\phi_2(y)=\exp\Bigl(\frac{i}{2}\,
\theta^{\mu\nu}\partial_\mu^x\partial_\nu^y\Bigr)
\phi_1(x)\phi_2(y)\vert_{x=y}
\label{3.2}
\end{equation}

\noindent
and $\theta^{\mu\nu}$ is a constant antisymmetric matrix.

\medskip
We notice that if one discards the surface terms, the derivatives of
the Moyal product will not contribute for the action (\ref{3.1}).
Concerning to the equation of motion, this can be done without further
considerations. However for the momenta we observe that one situation
and another are completely different. Let us then follow a similar
procedure to the previous section and make a variation of the action
(\ref{3.1}) by just keeping fixed the extreme in $t_0$. We obtain

\begin{eqnarray}
\delta S&=&\frac{1}{2}\int_{t_0}^tdt\int d^3\vec x\,
\Bigl(\partial_\mu\delta\phi\star\partial^\mu\phi
+\partial_\mu\phi\star\partial^\mu\delta\phi\Bigr)
\nonumber\\
&=&\frac{1}{2}\int_{t_0}^tdt\int d^3\vec x\,
\partial^\mu(\delta\phi\star\partial_\mu\phi
+\partial_\mu\phi\star\delta\phi)
\nonumber\\
&=&\frac{1}{2}\int d^3\vec x\,
(\delta\phi\star\dot\phi+\dot\phi\star\delta\phi)
\label{3.3}
\end{eqnarray}

\noindent
where it was used the on-shell condition. The momenta shall be
obtained from de development of (\ref{3.3}). Using the definition of
the Moyal product we have

\begin{eqnarray}
\delta S&=&\int d^3\vec x\,\Bigl[\dot\phi\,\delta\phi
\nonumber\\
&&+\frac{1}{2}\Bigl(\frac{i}{2}\Bigr)^2
\theta^{\mu\nu}\theta^{\alpha\beta}
\partial_\mu\partial_\alpha\dot\phi\,
\partial_\nu\partial_\beta\,\delta\phi
\nonumber\\
&&+\frac{1}{4!}\Bigl(\frac{i}{2}\Bigr)^4
\theta^{\mu\nu}\theta^{\alpha\beta}
\theta^{\rho\gamma}\theta^{\xi\eta}
\partial_\mu\partial_\alpha\partial_\rho\partial_\xi\dot\phi\,
\partial_\nu\partial_\beta\partial_\gamma\partial_\eta\,\delta\phi
\nonumber\\
&&+\cdots\Bigr]
\label{3.4}
\end{eqnarray}

\noindent
We observe that odd terms in $\theta^{\mu\nu}$ were canceled in
the expression above. This was so because of the symmetric terms that
appear in the first step of (\ref{3.3}), actually necessary in the
noncommutative case. In addition, due to the integration over
$d^3\vec x$ only terms in $\theta^{0i}$ will survive in the Moyal
product
\footnote{It was pointed out by Gomis and Mehen \cite{Gomis},
that the $\theta^{0i}$ should be taken zero in the vertex terms in
order to avoid causality and unitarity problems. However, the role
played by these terms in the free case is not the same.}.
For the quadratic term in $\theta^{\mu\nu}$ we obtain

\end{multicols}
\renewcommand{\theequation}{3.\arabic{equation}}
\Lrule

\begin{equation}
\int d^3\vec x\,
\theta^{\mu\nu}\theta^{\alpha\beta}
\partial_\mu\partial_\alpha\dot\phi\,
\partial_\nu\partial_\beta\delta\phi
=\int d^3\vec x\,
\bigl(\bar\partial^2\!\stackrel{...}{\phi}\!\delta\phi
+2\,\bar\partial^2\ddot\phi\,\delta\dot\phi
+\bar\partial^2\dot\phi\,\delta\ddot\phi\bigr)
\label{3.5}
\end{equation}

\noindent
where we have used the short notation $\bar\partial=\theta^{0i}
\partial_i$. Similarly, for the next term of the expression
(\ref{3.4}) we have

\begin{equation}
\int d^3\vec x\,\theta^{\mu\nu}\theta^{\alpha\beta}
\theta^{\rho\gamma}\theta^{\xi\eta}
\partial_\mu\partial_\alpha\partial_\rho\partial_\xi\dot\phi\,
\partial_\nu\partial_\beta\partial_\gamma\partial_\eta\delta\phi
=\int d^3\vec x\,
\bigl(\bar\partial^4\!\!\stackrel{(v)}{\phi}\!\delta\phi
+4\,\bar\partial^4\!\!\stackrel{(iv)}{\phi}\!\delta\dot\phi
+6\,\bar\partial^4\!\stackrel{...}{\phi}\!\delta\ddot\phi
+4\,\bar\partial^4\ddot\phi\,\delta\!\stackrel{...}{\phi}
+\bar\partial^4\dot\phi\,\delta\!\!\stackrel{(iv)}{\phi}\bigr)
\label{3.6}
\end{equation}

\noindent
where $\stackrel{(n)}{\phi}$ means $n$-time derivative over $\phi$. We
observe that the general rule to obtain other terms can be inferred.
Introducing these results into the initial Eq. (\ref{3.4}), grouping
the coefficients of $\delta\phi$, $\delta\dot\phi$, $\delta\ddot\phi$,
etc. and writing each term in a more convenient way, we have

\begin{eqnarray}
\delta S&=&\int d^3\vec x\,\biggl\{\biggl[
{0\choose0}\dot\phi
+{2\choose0}\frac{1}{2!}
\bigl(\frac{i}{2}\bar\partial\bigr)^2\stackrel{...}{\phi}
+{4\choose0}\frac{1}{4!}
\bigl(\frac{i}{2}\bar\partial\bigr)^4\stackrel{(v)}{\phi}
+\cdots\biggr]\,\delta\phi
\nonumber\\
&&\phantom{\int d^3\vec x\,\,}
+\biggl[{2\choose1}\frac{1}{2!}
\bigl(\frac{i}{2}\bar\partial\bigr)^2\ddot\phi
+{4\choose1}\frac{1}{4!}
\bigl(\frac{i}{2}\bar\partial\bigr)^4\stackrel{(iv)}{\phi}
+\cdots\biggr]\,\delta\dot\phi
\nonumber\\
&&\phantom{\int d^3\vec x\,\,}
+\biggl[{2\choose2}\frac{1}{2!}
\bigl(\frac{i}{2}\bar\partial\bigr)^2\dot\phi
+{4\choose2}\frac{1}{4!}
\bigl(\frac{i}{2}\bar\partial\bigr)^4\stackrel{...}{\phi}
+\cdots\biggr]\,\delta\ddot\phi
\nonumber\\
&&\phantom{\int d^3\vec x\,\,}
+\biggl[{4\choose3}\frac{1}{4!}
\bigl(\frac{i}{2}\bar\partial\bigr)^4\ddot\phi
+{6\choose3}\frac{1}{6!}
\bigl(\frac{i}{2}\bar\partial\bigr)^6\stackrel{(iv)}{\phi}
+\cdots\biggr]\,\delta\stackrel{...}{\phi}
+\cdots\biggr\}
\label{3.7}
\end{eqnarray}

\Rrule
\begin{multicols}{2}

\noindent
where

\begin{equation}
{p\choose n}=\frac{p!}{n!(p-n)!}
\label{3.8}
\end{equation}

\noindent
We can rewrite this relation in a compact form as

\begin{eqnarray}
&&\delta S=\int d^3\vec x\sum_{p,n=0}^\infty\biggl[
\frac{\bigl(\frac{i}{2}\bar\partial\bigr)^{2p}\,
\stackrel{(2p-2n+1)}{\phi}}{(2n)!(2p-2n)!}
\delta\!\stackrel{(2n)}{\phi}
\nonumber\\
&&\phantom{\delta S=}
+\frac{\bigl(\frac{i}{2}\bar\partial\bigr)^{2p+2}\,
\stackrel{(2p-2n+2)}{\phi}}{(2n+1)!(2p-2n+1)!}
\delta\!\stackrel{(2n+1)}{\phi}\biggr]
\label{3.9}
\end{eqnarray}

\noindent
One cannot still infer the momenta from the expression above because
$\delta\!\!\stackrel{(n)}{\phi}$ are not all independent. In fact, by
virtue of the equation of motion we have, for example,
$\delta\ddot\phi= \nabla^2\delta\phi$, $\delta\!\!\stackrel{...}\phi=
\nabla^2\delta\dot\phi$, and so on. Using these on-shell conditions
\cite{Landau} in expression (\ref{3.9}) and rewriting it in terms of
even and odd $n$, it is then finally possible to identify the
independent canonical momenta $\pi$ and $\pi^{(1)}$, conjugate
respectively to $\phi$ and $\dot\phi$,

\begin{eqnarray}
\pi&=&\sum_{p,n=0}^\infty
\frac{\bigl(\frac{i}{2}\bar\partial\sqrt{\nabla^2}\bigr)^{2p+2n}
\,\dot\phi}
{(2p)!(2n)!}
\label{3.10}\\
\pi^{(1)}&=&\sum_{p,n=0}^\infty
\frac{\bigl(\frac{i}{2}\bar\partial\sqrt{\nabla^2}\bigr)^{2p+2n+2}
\,\phi}
{(2p+1)!(2n+1)!}
\label{3.11}
\end{eqnarray}

\noindent
An interesting point is that a careful analysis of the expressions
above  permit us see that they can be cast in a closed form. We just
write down the final result

\begin{eqnarray}
\pi&=&\frac{1}{2}
\bigl[1+\cosh(i\bar\partial\sqrt{\nabla^2})\bigr]\dot\phi
\label{3.12}\\
\pi^{(1)}&=&\frac{1}{2}
\bigl[-1+\cosh(i\bar\partial\sqrt{\nabla^2})\bigr]\phi
\label{3.13}
\end{eqnarray}

The commutators cannot be directly obtained from the fundamental
Poisson brackets

\begin{eqnarray}
&&\{\phi(\vec x,t),\pi(\vec y,t)\}=\delta(\vec x-\vec y)
\label{3.14}\\
&&\{\dot\phi(\vec x,t),\pi^{(1)}(\vec y,t)\}
=\delta(\vec x-\vec y)
\label{3.15}
\end{eqnarray}

\noindent
because both relations (\ref{3.12}) and (\ref{3.13}) are constraints.
The calculation of the Dirac brackets can be done in a direct way (see
Appendix A). The most relevant bracket for the obtainment of the
propagator and the remaining of the quantization procedure is

\begin{equation}
\{\phi(\vec x,t),\dot\phi(\vec y,t)\}_D=\delta (\vec x-\vec y)
\label{3.16}
\end{equation}

\noindent
what means that the canonical quantization, even starting from the
nontrivial momentum expressions (\ref{3.10}) and (\ref{3.11}), or
(\ref{3.12}) and (\ref{3.13}), leads to same result of the
corresponding free commutative theory.

\section{Complex scalar fields}
\renewcommand{\theequation}{4.\arabic{equation}}
\setcounter{equation}{0}

We have seen in the previous analysis that there was a cancelation of
odd terms in $\theta^{\mu\nu}$ in the $\delta S$ on-shell variation
given by (\ref{3.3}). This was so due to the symmetry of the real
scalar fields in the action (\ref{3.1}). In this section we are going
to consider complex scalar fields where this symmetry does not exist
and consequently that cancelation does not occur.

\medskip
The noncommutative action for complex fields reads

\begin{equation}
S=\int d^4x\,\partial_\mu\phi^\ast\star\partial^\mu\phi
\label{4.1}
\end{equation}

\noindent
We could have also written here a symmetric quantity by adding a term
with $\partial_\mu\phi\star\partial^\mu\phi^\ast$ into the Lagrangian
of the action (\ref{4.1}). We are going to see that this is
nonetheless necessary because the action with
$\partial_\mu\phi\star\partial^\mu\phi^\ast$, even though having
different momenta expressions, leads to the same quantum result of the
one given by (\ref{4.1}). What is important to notice is that the
Lagrangian of the action (\ref{4.1}) does not have any problem related
to hermiticity, i.e., $(\partial_\mu\phi^\ast\star\partial^\mu\phi)
^\ast$ = $\partial_\mu\phi^\ast\star\partial^\mu\phi$, and discarding
boundary terms, the action (\ref{4.1}) leads to the usual free case
$S=\int d^4x\,\partial_\mu\phi^\ast\partial^\mu\phi$.

\medskip
Following the same steps as those of the previous section, we get

\begin{equation}
\delta S=\int d^3\vec x\,\bigl(\delta\phi^\ast\star\dot\phi
+\dot\phi^\ast\star\delta\phi\bigr)
\label{4.2}
\end{equation}

\noindent
whose a similar development permit us to obtain the momenta

\begin{eqnarray}
\pi&=&\sum_{p,n=0}^\infty
\frac{1}{p!(2n)!}
\bigl(-\frac{i}{2}\bar\partial\sqrt{\nabla^2}\bigr)^{p+2n}\,
\dot\phi^\ast
\label{4.3}\\
\pi^{(1)}&=&\sum_{p,n=0}^\infty
\frac{1}{p!(2n+1)!}
\bigl(-\frac{i}{2}\bar\partial\sqrt{\nabla^2}\bigr)^{p+2n+1}\,
\phi^\ast
\label{4.4}\\
\pi^\ast&=&\sum_{p,n=0}^\infty
\frac{1}{p!(2n)!}
\bigl(\frac{i}{2}\bar\partial\sqrt{\nabla^2}\bigr)^{p+2n}\,\phi
\label{4.5}\\
\pi^{(1)\ast}&=&\sum_{p,n=0}^\infty
\frac{1}{p!(2n+1)!}
\bigl(\frac{i}{2}\bar\partial\bigr)^{p+2n+1}\,\phi
\label{4.6}
\end{eqnarray}

\noindent
respectively conjugate to $\phi$, $\dot\phi$, $\phi^\ast$, and
$\dot\phi^\ast$. Also here, these sums lead to closed expressions


\begin{eqnarray}
\pi&=&\frac{1}{2}\bigl[
1+\cosh(i\bar\partial\sqrt{\nabla^2})
\bigr]\,\dot\phi^\ast
\nonumber\\
&&\phantom{\frac{1}{2}\bigl[1}
-\frac{1}{2}\sinh(i\bar\partial\sqrt{\nabla^2})\,
\sqrt{\nabla^2}\,\phi^\ast
\label{4.7}\\
\pi^{(1)}&=&\frac{1}{2}\bigl[
-1+\cosh(i\bar\partial\sqrt{\nabla^2})
\bigr]\,\phi^\ast
\nonumber\\
&&\phantom{\frac{1}{2}\bigl[1}
-\frac{1}{2}\sinh(i\bar\partial\sqrt{\nabla^2})\,
\frac{1}{\sqrt{\nabla^2}}\,\dot\phi^\ast
\label{4.8}\\
\pi^\ast&=&\frac{1}{2}\bigl[
1+\cosh(i\bar\partial\sqrt{\nabla^2})
\bigr]\,\dot\phi
\nonumber\\
&&\phantom{\frac{1}{2}\bigl[1}
+\frac{1}{2}\sinh(i\bar\partial\sqrt{\nabla^2})\,
\sqrt{\nabla^2}\,\phi
\label{4.9}\\
\pi^{(1)}&=&\frac{1}{2}\bigl[
-1+\cosh(i\bar\partial\sqrt{\nabla^2})
\bigr]\,\phi
\nonumber\\
&&\phantom{\frac{1}{2}\bigl[1}
+\frac{1}{2}\sinh(i\bar\partial\sqrt{\nabla^2})\,
\frac{1}{\sqrt{\nabla^2}}\,\dot\phi
\label{4.10}
\end{eqnarray}

\noindent
where the $sinh$-operators come from the odd terms in $\theta^{\mu\nu}
$. All the relations above are constraints. The calculation of the
Dirac brackets is a kind of direct algebraic work (see Appendix B).
The important point is that the brackets

\begin{eqnarray}
\{\phi(\vec x,t),\dot\phi^\ast(\vec y,t)\}_D
&=&\delta(\vec x-\vec y)
\nonumber\\
\{\phi^\ast(\vec x,t),\dot\phi(\vec y,t)\}_D
&=&\delta(\vec x-\vec y)
\label{4.11}
\end{eqnarray}

\noindent
are obtained, which means that the canonical quantization is correctly
achieved.

\medskip
To conclude this section, let us mention that we could have started
from

\begin{equation}
\tilde S=\int d^4x\,\partial_\mu\phi\star\partial^\mu\phi^\ast
\label{4.12}
\end{equation}

\noindent
instead of the action (\ref{4.1}). Considering the on-shell variation
of $\tilde S$ and keeping one of the extreme times fixed we have

\begin{equation}
\delta\tilde S=\int d^3\vec x\,\bigl(\delta\phi\star\dot\phi^\ast
+\dot\phi\star\delta\phi^\ast\bigr)
\label{4.13}
\end{equation}

\noindent
which leads to expressions for the momenta similar to
(\ref{4.7})-(\ref{4.10}) with a changing in the sign of the $sinh$-
terms. Even though the expressions for the momenta are not equivalent
in the two cases, we can trivially show that the constrained canonical
procedure leads to the same Dirac brackets given by (\ref{4.11}).

\section{Conclusion}
We have studied the free noncommutative scalar theory by using the
constrained canonical formalism in the appropriate form for dealing
with higher order derivative theories. This means that we have
considered the momenta as defined as the coefficients of $\delta\phi$
and $\delta\dot\phi$ calculated on the hipersurface orthogonal to
the time direction. We have shown that the evolved expressions coming
from the momenta definitions can be summed in a closed way making it
possible to be harmoniously applied in the Dirac constrained
formalism. We have also considered the complex scalar fields, where
the momentum expressions are still more evolved.

\medskip
These examples naturally obtained from noncommutative theories make
possible to verify the consistency of the constrained canonical
quantization procedure involving higher derivatives which is in some
sense a controversial subject in the literature.

\vspace{1cm}
\noindent
{\bf Acknowledgment:}
This work is supported in part by Conselho Nacional de Desenvolvimento
Cient\'{\i} fico e Tecnol\'ogico - CNPq (Brazilian Research agency)
with the support of PRONEX 66.2002/1998-9.

\section*{Appendix A}
\section*{Dirac brackets for the real scalar case}
\renewcommand{\theequation}{A.\arabic{equation}}
\setcounter{equation}{0}

Let us denote constraints (\ref{3.12}) and (\ref{3.13}) in a
simplified notation like

\begin{eqnarray}
T^0&=&\pi+K\,\dot\phi
\label{A.1}\\
T^1&=&\pi^{(1)}+L\,\phi
\label{A.2}
\end{eqnarray}

\noindent
where $K$ and $L$ are the operators

\begin{eqnarray}
K&=&-\,\frac{1}{2}
\bigl[1+\cosh(i\bar\partial\sqrt{\nabla^2})\bigr]
\label{A.3}\\
L&=&\frac{1}{2}
\bigl[1-\cosh(i\bar\partial\sqrt{\nabla^2})\bigr]
\label{A.4}
\end{eqnarray}

\noindent
Using the fundamental Poisson brackets given by (\ref{3.14}) and
(\ref{3.15}) we have

\begin{eqnarray}
\{T^0(\vec x,t),T^1(\vec y,t)\}&=&-\,L_y\,\delta(\vec x-\vec y)
+K_x\,\delta(\vec x-\vec y)
\nonumber\\
&=&-\,\delta(\vec x-\vec y)
\nonumber\\
&=&-\,\{T^1(\vec x,t),T^0(\vec y,t)\}
\label{A.5}
\end{eqnarray}

\noindent
where in the last step there was a providential cancelation of the
even operators $\cosh(i\bar\partial\sqrt{\nabla^2})$ acting on
$\delta(\vec x-\vec y)$. Since the remaining Poisson brackets of the
constraints are zero, we have that the corresponding Poisson brackets
matrix is given by

\begin{equation}
M=\left(\begin{array}{cc}
0&-\,1\\
1&0
\end{array}\right)\,\delta(\vec x-\vec y)
\label{A.6}
\end{equation}

\noindent
whose inverse is directly obtained, as is also the Dirac brackets
(\ref{3.16}).

\section*{Appendix B}
\section*{Dirac brackets for the complex case}
\renewcommand{\theequation}{B.\arabic{equation}}
\setcounter{equation}{0}

Let us denote the corresponding constraints by

\begin{eqnarray}
T^0&=&\pi+K_1\,\dot\phi^\ast+K_2\,\phi^\ast
\nonumber\\
T^1&=&\pi^{(1)}+L_1\,\dot\phi^\ast+L_2\,\phi^\ast
\nonumber\\
T^2&=&\pi^\ast+K_1\,\dot\phi-K_2\,\phi
\nonumber\\
T^0&=&\pi^{(1)\ast}-L_1\,\dot\phi+L_2\,\phi
\label{B.1}
\end{eqnarray}

\noindent
where $K_1$, $K_2$, $L_1$, and $L_2$ are short notations for the
operators

\begin{eqnarray}
K_1&=&-\,\frac{1}{2}
\bigl[1+\cosh(i\bar\partial\sqrt{\nabla^2})\bigr]
\nonumber\\
K_2&=&\frac{1}{2}\,
\sinh(i\bar\partial\sqrt{\nabla^2})\sqrt{\nabla^2}
\nonumber\\
L_1&=&\frac{1}{2}\,
\sinh(i\bar\partial\sqrt{\nabla^2})\frac{1}{\sqrt{\nabla^2}}
\nonumber\\
L_2&=&\frac{1}{2}
\bigl[1-\cosh(i\bar\partial\sqrt{\nabla^2})\bigr]
\label{B.2}
\end{eqnarray}

\noindent
The Poisson brackets for these constraints are

\begin{eqnarray}
\{T^0(\vec x,t),T^2(\vec y,t)\}
&=&\bigl(K_{2y}+K_{2x}\bigr)\,\delta(\vec x-\vec y)=0
\nonumber\\
\{T^0(\vec x,t),T^3(\vec y,t)\}
&=&\bigl(-L_{2y}+K_{1x}\bigr)\,\delta(\vec x-\vec y)
\nonumber\\
&=&-\,\delta(\vec x-\vec y)
\nonumber\\
\{T^1(\vec x,t),T^2(\vec y,t)\}
&=&\bigl(K_{1y}+L_{2x}\bigr)\,\delta(\vec x-\vec y)
\nonumber\\
&=&\delta(\vec x-\vec y)
\nonumber\\
\{T^1(\vec x,t),T^3(\vec y,t)\}
&=&\bigl(L_{1y}+L_{1x}\bigr)\,\delta(\vec x-\vec y)=0
\label{B.3}
\end{eqnarray}

\noindent
The remaining brackets are trivially zero. It is interesting to
observe the harmonious cancelation among the different operators
acting on the delta function. Now, we can easily construct the matrix
of the Poisson brackets of the constraints and calculate the relevant
Dirac brackets given by (\ref{4.11}).

\end{multicols}
\end{document}